\documentclass[preprint,showpacs,preprintnumbers,amsmath,amssymb]{revtex4}

\usepackage{graphicx}
\usepackage{dcolumn}
\usepackage{bm}

\preprint{APS/123-QED}

\begin{document}

\title{
Dephasing of coupled qubit system during gate operations 
due to background 
charge fluctuations
}


\author{Toshifumi Itakura  
and Yasuhiro Tokura }

\affiliation{
   NTT Basic Research Laboratories, NTT corporation \\
  3-1, Morinosato Wakamiya, Atsugi-shi,
  Kanagawa Pref., 243-0198 Japan \\
    Tel : +81-46-240-3938
    Fax : +81-46-240-4726  \\
    tokura@will.brl.ntt.co.jp \\
}

\date{\today}

\begin{abstract}
A quantum computer that can 
be constructed
  based on a superconducting nanocircuits has previously been proposed.
We examine the effect of background charge fluctuations 
on a coupled Josephson charge qubit system used in such a computer. 
In previous work, the background charge fluctuations were found
 to be an important dephasing channel for a single Josephson qubit.
We investigate the effect of fluctuations in the bias at
the charge degeneracy point of a Josephson charge qubit system.
Evaluated quantities are gate fidelity and diagonal elements
 of the qubit's density matrix.
The fluctuation leads to gate error, however quantum gate operation 
  becomes more accurate with increasing interaction between qubit systems.

\end{abstract}


\maketitle

\section{Introduction}

Among the various proposals for quantum computation,
 quantum bits (qubits) in solid state materials,
           such as superconducting Josephson junctions
\cite{Nakamura}
           and quantum dots
\cite{Hayashi,Tanamoto,Loss},
           have the advantage of  scalability.
Proposals to implement a quantum computer using superconducting nanocircuits
    are proving to be very promising
    \cite{Makhlin,Makhlin_RMP,Averin,Mooij,Falci},
    and several experiments have already highlighted the quantum properties
    of these devices
    \cite{Bouchiat,Friedman,Aassime}.
Such a coherent-two-level system constitutes a qubit and
 the quantum computation can be carried out as the
        unitary operation functioning on the multiple qubit system.
Essentially, this quantum coherence must be maintained during computation.
However, it is difficult to avoid dephasing  
     due to the system's interaction with
     its external environment.
The dephasing is characterized by a dephasing time of $T_2$.
Various environments can cause dephasing.

Background charge fluctuations (BCFs) 
         have been observed in diverse kinds of  systems
\cite{Devoret,Zorin,Martinis,Lyon}.
In nanoscale systems, BCFs are
     electrostatic potential fluctuations arising
     due to the dynamics of an electron, or a hole,
     on a charge trap.     
In particular, the charges in charge traps 
     fluctuate
     with the  Lorentzian spectrum form,
     which is 	
     called
     random telegraph noise
     in the time domain
\cite{Lyon,Fujisawa_BC}.
The random 
      distribution of the positions of such dynamical charge traps
       and their time constants  
     leads to  BCFs or 1/f noise
\cite{BC}.
In solid-state charge qubits,
    these BCFs
    result in  a  dynamical electrostatic disturbance and
    hence, dephasing.
It should be noted that this dephasing process does not mean 
        the qubit 
        being entangled with the environment,
        but rather,
        that  
        the stochastic evolution of an external classical field
        is suppressing the density matrix elements 
        of the qubit after  averaging out over statistically distributed
         samples.

It has been shown that BCFs are important dephasing channels 
  for a single Josephson charge qubit system
\cite{Itakura_Tokura_PRB,Itakura_Tokura,Nakamura_CE,Fazio,Shnirman}.
In the present study, we investigate the effect of BCFs on the two-qubit-gate
    operation.
To construct a controllable quantum computer, one requires 
    the suppression of dephasing and
    accurate  universal quantum gate which  consists of
    single qubit operations and two-qubit operations is required.
Therefore, to address these manipulations, we examine 
     the dephasing of a coupled qubit system,	
     which is an experimentally current topic.
There is a lot of interest in understanding what causes
     dephasing and its role in these systems.               
In Sec. II, we discuss  the dephasing in a Josephson charge qubit system.
Sec III is a brief conclusion.        
Similar subjects are also discussed in terms of decoherence-free subspace 
   in Ref. \cite{Nakano}.

\section{Coupled Charge qubit system}

The system under consideration is a pair of Cooper pair boxes
     \cite{Makhlin}.
Under appropriate conditions (charging energy $ E_{C_{1,2}}$ and 
the Josephson coupling $E_{J_{1,2}}$ are
 much larger than  and temperatures $k_B T \ll E_{J_{1,2}},E_{C_{1,2}}$)
     only two charge states in each box are important,
     and the Hamiltonian of the pair of qubits  $H_{qb}$ reads      
\begin{equation}
  H_{qb} = \frac{E_{J1}}{2} \sigma_x^1   + \frac{E_{J2}}{2} \sigma_x^2
            + \frac{\delta E_{C1}}{2} \sigma_z^1 
            + \frac{\delta E_{C2}}{2} \sigma_z^2
            + \frac{E_{m}}{4}  \sigma_z^1 \sigma_z^2,  
\end{equation}
where  the charge bases $\{| 0 \rangle ,|1 \rangle\}$ 
are expressed using Pauli matrices.
We chose the charge degeneracy point $\delta E_{C_{1,2}} =0$ 
 except for our last result,
 where $\delta E_{C_{i}} = E_{C_{i}} (1 - C_x^i V_x^i/e)$, $(i=1,2)$
 and $C_x^i$ and $V_x^i$ are capacitance 
 and gate bias of i-th Cooper pair box.       
The environment is a set of BCFs electrostatistically coupled to the qubits
\cite{Itakura_Tokura_PRB,Itakura_Tokura,Nakamura_CE,Fazio,Shnirman,Galperin,Nazarov},
\begin{equation}             
  H_{qb} =  
  \frac{ \hbar J }{2} ( \sigma_z^1 + \sigma_z^2 ) 
  ( d^{\dagger} d - \frac{1}{2} ),
\end{equation}
        where $d^{\dagger}$ and $d$ are 
        the electron creation and annihilation operators
        of a charge trap,
        and the coupling with the qubit is such that each BCF produces
        a bistable extra bias 
         $\hbar J$.
Because the qubit Hamiltonian consists of $E_{J_{1,2}}$ and $E_{m}$,
        the dephasing is accompanied by dissipation.
It should be noted that we evaluated a {\it collective}  environment.
The two Cooper pair boxes feel the same fluctuations in our model.

Using the environment variable  
        $X (t)  (= \langle d^{\dagger} (t) d (t) \rangle_r - 1/2 )$,
        where $\langle A(t) \rangle_r$ is a trace 
         of the operator $A(t)$ 
        about the electron reservoir of the charge trap, 
        we rewrite
        the perturbation Hamiltonian 
        in terms of the  Pauli matrix as
\begin{equation}
{\cal H}_1 \equiv \frac{\hbar J}{2} (\sigma_z^1 + \sigma_z^2 ) X (t)
= J X (t)  V_1,
\end{equation}
where we denote that the charge trap is strongly coupled with
            its charge reservoir
            and the time evolution of $ X(t)$
            is a Poisson process (BCF).
We assume $\langle X(t_1) X(t_2) \rangle =  e^{-|t_1 - t_2|/\tau}$,
  which corresponds to dephasing due to a single trap
where $\langle \rangle$ denotes the ensemble average       
  and $\tau$ is the time constant of a BCF.  
In the interaction representation,
  $U_0 (t) = e^{-\frac{i}{\hbar} H_{qb} t}$,
$U_1 (t) = e^{-\frac{i}{\hbar} \int_0^t \overline{H}_1 (t') dt'}$
  and
  $ \overline{V}_1 (t) = U_0 (t) V_1 U_0^{\dagger} (t)$.
The ensemble averaged density matrix $\rho(t)$ at time $t$
  in a second-order perturbation
  approximation is,
$ \langle \rho(t) \rangle = U_0 (t) \rho(0) U_0^\dagger (t)
   - J^2 \int_0^t dt_1 \int_0^{t_1} dt_2 
  U_0 (t) e^{\frac{-t_1+t_2}{\tau}} [
  \overline{V}_1 (t_1) ,
   [ \overline{V}_1 (t_2), \rho (0) ]] U_0^{\dagger} (t). $
The gate fidelity is defined as
 ${\cal F}=   Tr (\rho_0 (t) \langle \rho (t) \rangle ) $,
 where $\rho_0 (t) = U_0 (t) \rho (0) U_0^\dagger (t) $.
In Fig. 1, we show the $E_J t$ dependence of $-\ln {\cal F}(t)$,
  where the initial density matrix is $|00 \rangle \langle 00|$.
For simplicity we set $E_{J1}=E_{J2}$.  
The solid lines denote $E_{m}/E_J$=20 and the dotted lines denote
   $E_{m}=0$.  
We choose parameters $J / E_J$=0.5
and  $E_J \tau= 10^{-1}, 10^{-2}, 10^{-3}$. 
At $t$=0, fidelity is 1, and it decreases with time.
In the short time regime $t \ll \frac{\hbar}{E_J} $,
the fidelities show $-\ln {\cal F} (t) \propto t^4$.
The lowest-order Gaussian behavior $(t^2)$ originating from the term like
 $ [ \sigma_z^1 + \sigma_z^2,
  [ \sigma_z^1 + \sigma_z^2 , \rho(0)]]$ is absent,
since we started from a diagonal qubits' density matrix.
As $E_J \tau$ increases, fidelity worsens 
\cite{Itakura_Tokura_PRB}.
The fidelity of $E_{m} \ne 0$ is larger than that of $E_{m}=0(>0)$.
The reason is as follows.
The fluctuation $V_1$ only induces transitions
   between the ground state and the 2nd excited state of $H_{qb}$,
  and between  the 2nd excited  and the 3rd excited state.
When $E_{m}$=0, the sum of the dephasing rates 
   for these two transitions is given by
   $2 J^2 \tau$.
As $E_{m}$ increases, the dephasing rate decreases down to $J^2 \tau$,  
   since the dephasing by the transition between the ground state and
   the 2nd excited state is gradually suppressed 
   \cite{Itakura_Tokura_PRB}.
Thus, the dephasing rate of $E_{m} = 0 $ is larger than that of $E_{m} \ne 0$.
Therefore, it is expected that interaction between the qubits leads to more
  reliable quantum gate operations.

Figure 2 shows the Positive Operator-Value Measurement (POVM)
 results where $I_1$ is the sum of
 the density matrix elements of
  $\langle 10 | \langle \rho (t)  \rangle | 10 \rangle$
  and $\langle 11 | \langle \rho (t) \rangle |  11 \rangle$,
 with parameters $J_C =1$ GHz, $\tau=0.1$ ns,
  $E_{J1}=13.4$ GHz,
 $E_{J2}=9.1$ GHz and $E_{m}=11.6$GHz,
 which are the same as those in
 the experiment in Ref.
 \cite{Nakamura_Nature}.
We compared the Fourier spectrum for $J=0$ and that of $J=1$GHz.
There are two peaks in Fourier spectrum with finite width 
  which correspond to  different $E_{J_{1,2}}$ of qubit system.
The spectrums also 
show that the   peak widths of the qubit system are larger for $|J| \ne 0$.
This signals that  the dephasing due to BCF occurs,
as we have shown in the analysis of gate fidelity.
For fidelity, ${\cal F}(t)$=1 
when $J=0$ and ${\cal F} (t)<1 $ when $|J| \ne 0$.

We also examined the time evolution of coherent transition  between
 $|00 \rangle \iff |10 \rangle$.
We choose $\delta E_{C_1}=0$, $\delta E_{C_2}$=152 GHz and $E_{m}=11.6$ GHz.
This corresponds to single qubit coherence oscillation.  
The coherent oscillation is robust against BCF
  compared with the case of two-qubit operations. 
This behavior is consistent with the experiment 	
Ref. \cite{Nakamura_Nature}.    
We speculate that this result depends on the way of 
 the environment coupling to qubits.

\section{Conclusion}

We examined the effect of BCF on the coupled qubit system during
  gate operation.
The fluctuation leads to gate error, however quantum gate operation 
  becomes more accurate with increasing  interaction between qubit systems
  when one start from $|00 \rangle \langle 00|$.

{\bf Acknowledgements} 
The authors thank A. Kawaguchi, T. Hayashi, 
T. Fujisawa, T. Tanamoto, S. Lloyd, and
Y. Hirayama
for their advice and stimulating discussions.
This work was partly supported by SORST-JST.

\begin{figure}
\center
\includegraphics[width=0.7\linewidth]{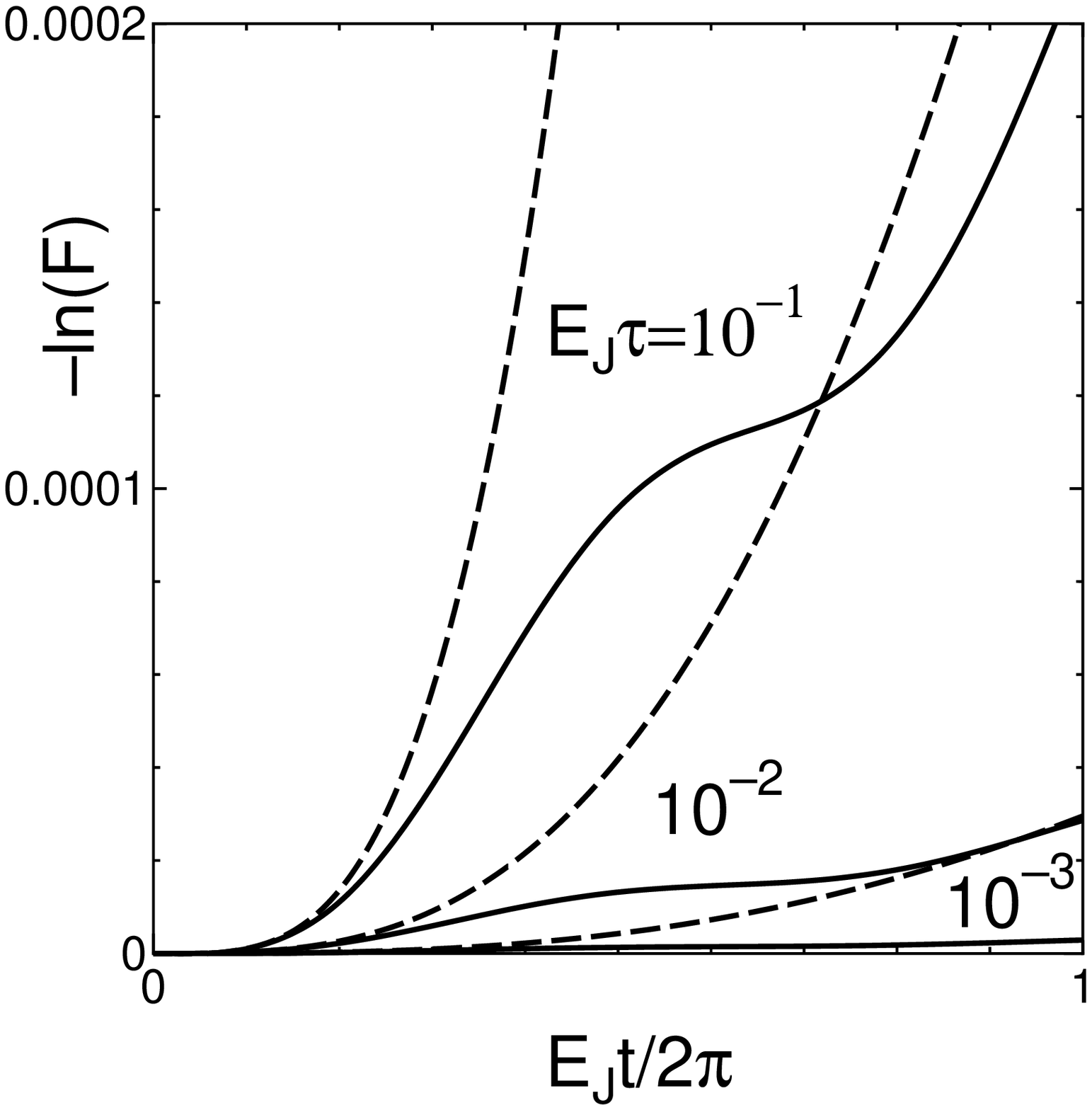}
\begin{minipage}[t]{6cm}
\caption{
Time dependence of gate fidelity of two-qubit operation.
}
\label{figure1}
\end{minipage}
\end{figure}
\begin{figure}
\center
\includegraphics[width=0.7\linewidth]{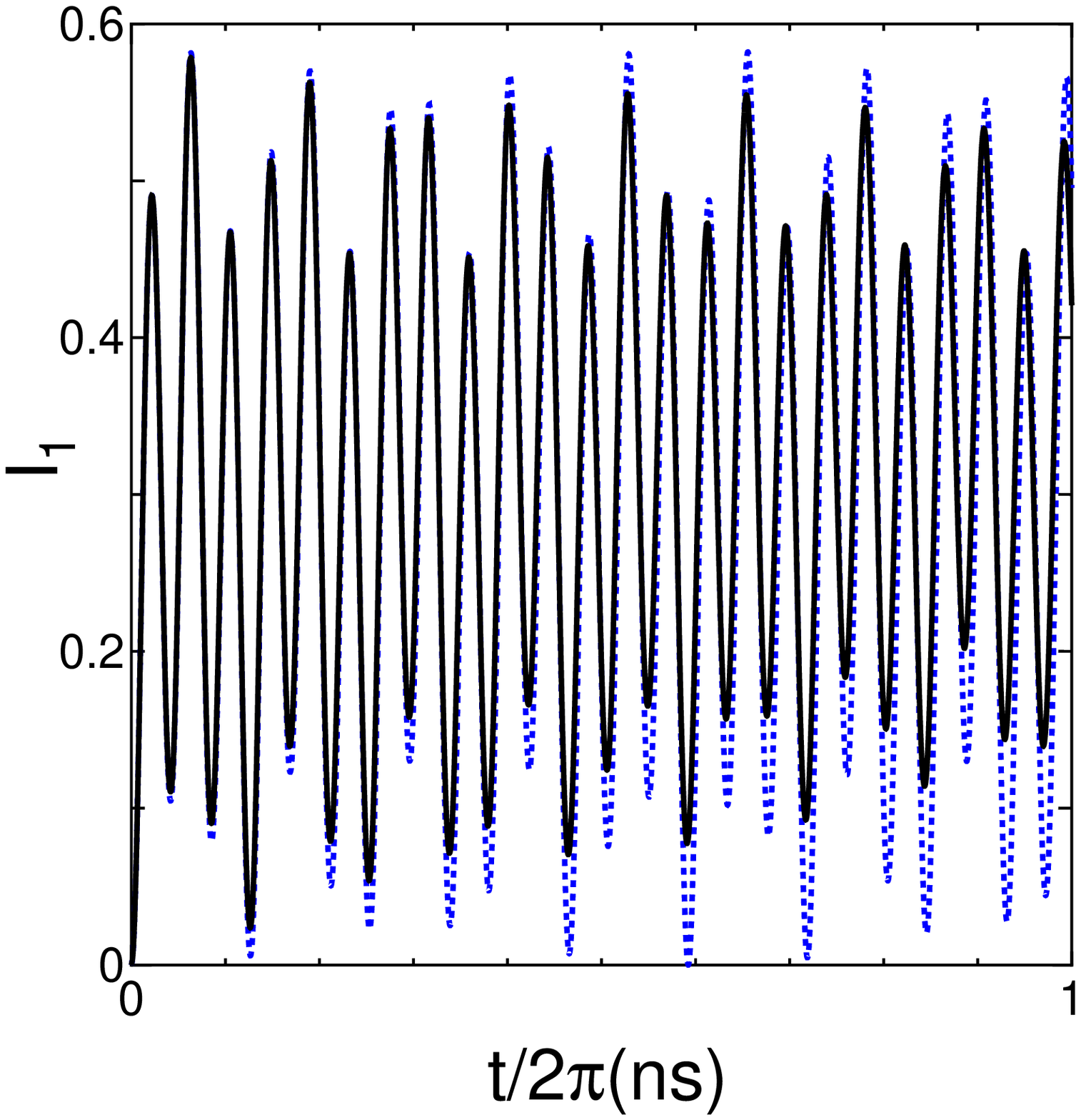}
\begin{minipage}[t]{6cm}
\caption{
Time dependence of the diagonal element of the qubit density matrix.
}
\label{figure2}
\end{minipage}
\end{figure}


\begin{references}
\bibitem{Nakamura}
   Y. Nakamura, Yu. A. Pashkin and J. S. Tsai:
  Nature {\bf 398}  (1999) 786.
\bibitem{Hayashi}
  T. Hayashi, T. Fujisawa, H-D. Cheong, Y-H. Jeong and Y. Hirayama:
  Phys. Rev. Lett. {\bf 91} (2003) 226804.
\bibitem{Tanamoto}
   T. Tanamoto:
   Phys. Rev. A {\bf 61}  (2000) 22305.
\bibitem{Loss}
   D. Loss and D. P. DiVincenzo:
   Phys. Rev. A {\bf 57}  (1998) 120.
\bibitem{Makhlin}
 Y. Makhlin, G. Sch\"on and Shniriman: Nature {\bf 398}  (1999) 305.
\bibitem{Makhlin_RMP}
Y. Makhlin, G. Sch\"on and A. Shnirman: Rev. Mod. Phys. {\bf 73}  (2001) 357.
\bibitem{Averin}
D. A. Averin: Solid State Commun. {\bf 105}  (1998) 659.
\bibitem{Mooij}
J. E. Mooij, T. P. Orland, L. Levitov, L. Tian, C. H. van der Wal
 and S. Lloyd: Science {\bf 285}  (1999) 1036.
\bibitem{Falci}
G. Falci, R. Fazio, G. M. Palma, J. Siewert and V. Verdral:
  Nature  {\bf 407}  (2000) 355.
\bibitem{Bouchiat}
 V. Bouchiat, D. Vion, P. Joyez, D. Esteve and M. H. Devoret:
  Phys. Scr. {\bf T76}  (1998) 165.
\bibitem{Friedman}
 J. R. Friedman, V. Patel, W. Chen, S. K. Tolpyo and J. E. Lukons:
  Nature  {\bf 406}  (2000) 43.
\bibitem{Aassime}
 A. Aassime, G. Johansson, G. Wendin, R. J. Schoellcopf and
 P. Delsing: Phys. Rev. Lett. {\bf 86}  (2000) 3376.
\bibitem{Devoret}
   P. L. Lafarge, P. Joyez, H. Pothier, A. Cleland, T. Holst,
   D. Esteve, C. Urbina and M. H. Devoret:
   C. R. Acad. Sci. Paris, {\bf 314}  (1992) 883.
\bibitem{Zorin}
    A. B. Zorin, F.-J. Ahlers, J. Niemeyer, T. Weimann, H. Wolf,
    V. A. Krupenin and S.V. Lotkhov:
    Phys. Rev. B {\bf 53}  (1996) 13682.    
\bibitem{Martinis}
   G. Zimmerli, T. M. Eiles, R. L. Kautz and J. M. Martinis:
   Appl. Phys. Lett. {\bf 61}  (1992) 13.
\bibitem{Lyon}
    C. Kurdak, C. J. Chen, D. C. Tsui, S. Parihar, S. Lyon and G. W. Weimann:
    Phys. Rev. B {\bf 56}  (1997) 9813.
\bibitem{Fujisawa_BC}  T. Fujisawa and Y. Hirayama:
  Appl. Phys. Lett {\bf 77}  (2000) 543.
\bibitem{BC}  P. Dutta and P. H. Horn:
  Rev. Mod. Phys. {\bf 53}  (1981) 497.
\bibitem{Itakura_Tokura_PRB}
T. Itakura and Y. Tokura: 
Phys. Rev. B {\bf 67} (2003) 195320.
\bibitem{Itakura_Tokura}
T. Itakura and Y. Tokura: J. Phys. Soc. Jpn. {\bf 72}  (2003) 2726.
\bibitem{Nakamura_CE}
        Y. Nakamura, Yu A. Pashkin, T. Yamamoto and J. S. Tsai:
        Phys. Rev. Lett. {\bf 88}  (2002) 047901.
\bibitem{Fazio}
 E. Paladino, L. Faoro, G. Falci and R. Fazio: 
  Phys. Rev. Lett. {\bf 88}  (2002) 228304.
\bibitem{Shnirman}
  A. Shnirman, Y. Makhlin and G. Sch\"on:
  Phys, Scr. {\bf T102}  (2002) 147.  
\bibitem{Nakano}
 H. Nakano and H. Takayanagi:
 {\it Proceedings of the 7th International Symposium on
 Foundation of Quantum Mechaincs in the Light of New Technology}
 ISQM-TOKO'01 28, World Scientific, Singapore (2002).
\bibitem{Galperin}
Y. M. Galperin and K. A. Chao: 
  Phys. Rev. B {\bf 52}  (1995) 12126. 
\bibitem{Nazarov}
  R. Bauernschmitt and Y. V. Nazarov:
  Phys. Rev. B {\bf 47}  (1992) 9997.
\bibitem{Nakamura_Nature}
 Y. A. Pashkin, T. Yamamoto, O. Astafive, Y. Nakamura, D. V. Averin
 and J. S. Tsai: Nature {\bf 421} 823 (2003).
\end{references}
\end{document}